\documentclass[aps,pre,twocolumn,floats,showpacs]{revtex4}
\usepackage{epsfig}
\usepackage{color}
\usepackage{bm}
\usepackage{latexsym}

\begin{document}
\newcommand{\hide}[1]{}
\newcommand{\tbox}[1]{\mbox{\tiny #1}}
\newcommand{\half}{\mbox{\small $\frac{1}{2}$}}
\newcommand{\sinc}{\mbox{sinc}}
\newcommand{\const}{\mbox{const}}
\newcommand{\trc}{\mbox{trace}}
\newcommand{\intt}{\int\!\!\!\!\int }
\newcommand{\ointt}{\int\!\!\!\!\int\!\!\!\!\!\circ\ }
\newcommand{\eexp}{\mbox{e}^}
\newcommand{\bra}{\left\langle}
\newcommand{\ket}{\right\rangle}
\newcommand{\EPS} {\mbox{\LARGE $\epsilon$}}
\newcommand{\ar}{\mathsf r}
\newcommand{\im}{\mbox{Im}}
\newcommand{\re}{\mbox{Re}}
\newcommand{\bmsf}[1]{\bm{\mathsf{#1}}}
\newcommand{\mpg}[2][1.0\hsize]{\begin{minipage}[b]{#1}{#2}\end{minipage}}

\title{Disorder to chaos transition in the conductance distribution of
corrugated waveguides}

\author{A. Alc\'azar-L\'opez and J. A. M\'endez-Berm\'udez}
\affiliation{Instituto de F\'{\i}sica, Benem\'erita Universidad Aut\'onoma de Puebla,
Apartado Postal J-48, Puebla 72570, Mexico}

\date{\today}

\begin{abstract}
We perform a detailed numerical study of the distribution of conductances $P(T)$
for quasi-one-dimensional corrugated waveguides as a function of the  
corrugation complexity (from rough to smooth). We verify the universality of $P(T)$ 
in both, the diffusive ($\bra T \ket> 1$) and the localized ($\bra T \ket\ll 1$) 
transport regimes. However, at the crossover regime ($\bra T \ket \sim 1$), we 
observe that $P(T)$ evolves from the surface-disorder to the bulk-disorder 
theoretical predictions for decreasing complexity in the waveguide 
boundaries. We explain this behavior as a transition from disorder to deterministic 
chaos; since, in the limit of smooth boundaries the corrugated waveguides are, 
effectively, linear chains of chaotic cavities.
\end{abstract}


\maketitle


\section{Introduction}

In studies of wave propagation through disordered wires two setups are mostly 
considered: bulk-disordered and surface-disordered waveguides.
In both cases it is possible to discern between diffusive (metallic) and localized 
(insulating) transport regimes by comparing the wire length $L$ with the mean-free 
path $\ell$ and the localization length $\xi$; i.e., wave diffusion takes place when 
$\ell\ll L\ll\xi$ while localization is observed for $\xi\ll L$.

From the analytical point of view both disorder setups have been successfully 
approached. On the one hand, transport through bulk-disordered waveguides is well 
described by the Fokker-Planck approach of Dorokhov, Mello, Pereyra, and Kumar 
(DMPK) \cite{D82,MPK88,M88,MS91,MC92,MK04};
and by the field-theoretic approach of Efetov and Larkin, which leads to a supersymmetric
nonlinear $\sigma$ model \cite{EL83,E83,MMZ94}. In fact, these two approaches were  
shown to be equivalent in Ref.~\cite{BF96}.
In addition, in Ref.~\cite{F03}, the distribution of conductances $P(T)$ in the full 
diffusive-to-localized crossover was derived, in the frame of the supersymmetric 
approach, for waveguides with broken time-reversal invariance \cite{note0b}.
On the other hand, transport through surface-disordered wires has been properly 
characterized by the Fokker-Planck approach developed by Froufe-Perez, Yepez, Mello, 
and Saenz (FYMS) \cite{FYMS07}. 
Other analytical approaches to transport through surface-disordered wires are also 
available in Refs.~\cite{MI1,MI2}.

Furthermore, for both setups, $P(T)$ evolves from a Gaussian shape (deep in the 
diffusive regime) to a log-normal shape (deep in the localized regime). However, 
at the crossover between diffusive and localized regimes, the form of $P(T)$ is highly 
non-trivial \cite{MK04,FYMS07,GTSN97,GS01,FGSMS02,GS05,GMW02,CFG05,PW98,M02}
and importantly depends on the type of disorder (bulk or surface). Moreover, the 
DMPK approach and the FYMS approach provide accurate predictions for $P(T)$ at the crossover regime 
for the corresponding setups of disorder \cite{MK04,FYMS07,GTSN97,GS01,FGSMS02,GS05,GMW02}.

In this paper we numerically study $P(T)$ for a model of quasi-one-dimensional 
surface-disordered waveguides with tunable corrugation complexity: from rough 
to smooth. Here, we concentrate on waveguides with time-reversal invariance. 
We define the corrugated surface of our disordered wire as a sum
of harmonics with random amplitudes. In the rough limit (large number of harmonics)
the waveguide effectively shows surface disorder; so that, it is equivalent to
the step-like corrugated waveguide model used in 
Refs.~\cite{GTSN97,GS01,FGSMS02,GS05}.
On the other hand, in the smooth limit (few harmonics) the waveguide can be 
considered as a linear chain of attached chaotic cavities. Interesting enough,
the transport properties of a waveguide constructed as a linear chain of chaotic 
cavities \cite{IWZ90,MMZ94,B07} are equivalent to those of a 
bulk-disordered wire \cite{BF96}. Then, by decreasing the corrugation complexity 
of our surface corrugated waveguide we expect to observe, at the 
diffusive-to-localized transition regime, a transition in the form 
of $P(T)$ from the surface-disorder FYMS to the bulk-disorder DMPK predictions.

The organization of this paper is as follows.
In the next section we define the waveguide model we use as well as the 
scattering setup.
In Sec. III, by extracting $\ell$ and $\xi$ from curves of the average resistance
and average logarithm of conductance as a function of $L$, respectively, we define 
the diffusive and localized transport regimes for our corrugated waveguides.
Then we study in detail the distribution of conductances $P(T)$ as a function of 
the corrugation complexity in both regimes, diffusive and localized, as well as at 
the crossover regime. Finally, Sec. IV is left for conclusions.

\section{Model}

\begin{figure}[t]
\includegraphics[width=8cm]{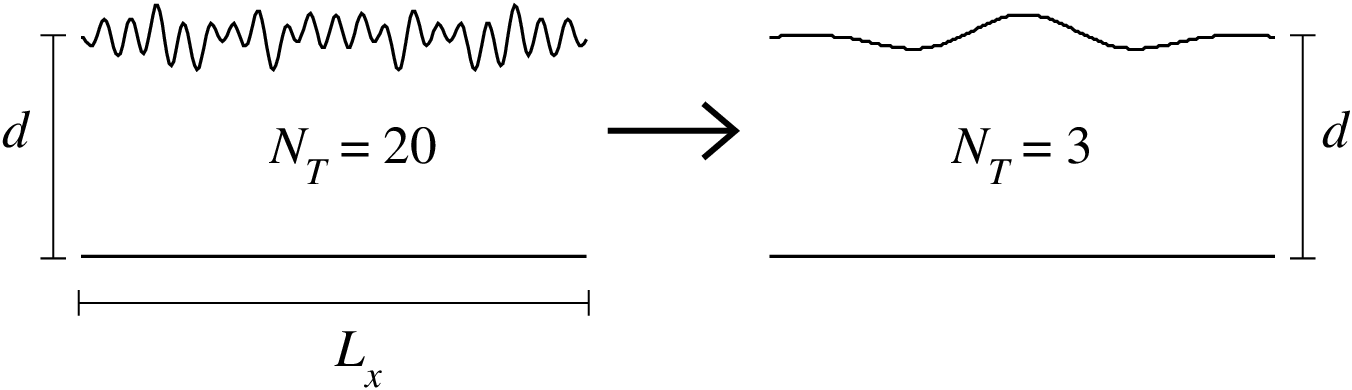}
\caption{Examples of cavities used to form corrugated waveguides. 
Here we show one realization of cavities with $N_T=20$ and $N_T=3$.
In our study we go from $N_T=18$ to $N_T=2$.}
\label{Fig1}
\end{figure}

The model we shall use in our study is a waveguide formed by attaching $L$ 
two-dimensional cavities. Each cavity of length $L_x$ is 
defined by two hard 
walls: one flat at $y=0$ and a corrugated one given by $y=d+\epsilon f(x)$. 
Here $d$ is the average width of the cavity and $\epsilon$ is the corrugation 
amplitude.
Since we are interested in studying the transport properties of waveguides
as a function of the complexity of the corrugated boundary we choose
\begin{equation}
f(x) = \sum_{n=1}^{N_T} A_n \cos\left( \frac{2\pi n}{L_x} x \right) \ ,
\end{equation}
where $A_n$ are random numbers drawn from a flat distribution in the range
$[-1,1]$. This form for $f(x)$ allows us to choose the desired degree of complexity 
of the corrugated waveguide boundary: from rough, $N_T\sim 20$, to smooth, 
$N_T\sim 1$. See Fig.~\ref{Fig1}. It is important to stress that once the 
parameters $L_x$, $d$, $\epsilon$, and $N_T$ are fixed we randomly generate 
(through different values of $A_n$) $L$ different cavities that we attach 
to form a non-periodic
waveguide \cite{note1,note1b}. To this end the minimal $N_T$ we use is 2.

\begin{figure*}[t]
\includegraphics[width=11cm]{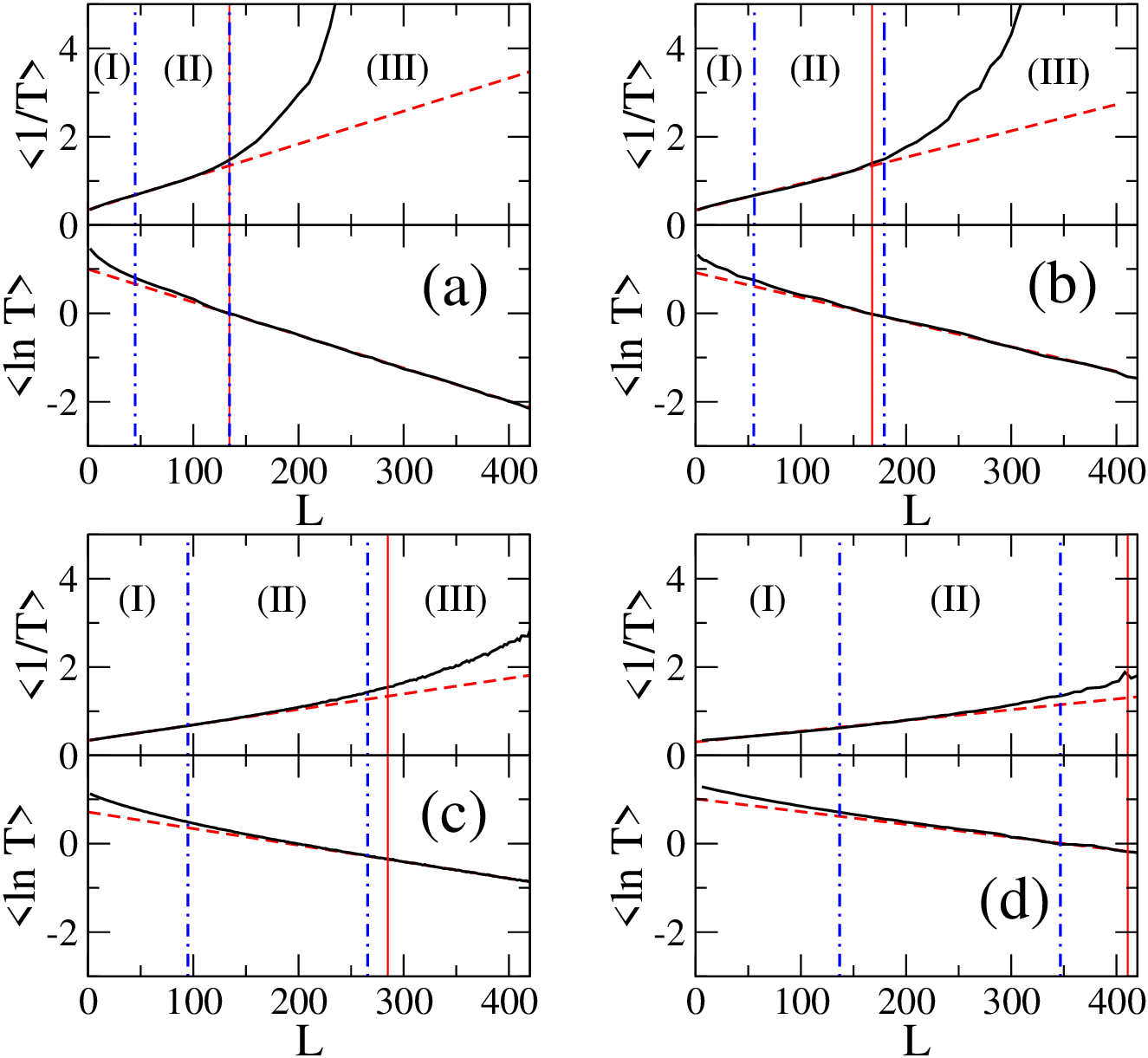}
\caption{(Color online) Average resistance $\langle 1/T \rangle$ (upper 
panels) and average logarithm of conductance $\left\langle \ln T \right\rangle$ 
(lower panels) as a function of the length $L$ of corrugated waveguides with 
(a) $N_T=18$, (b) 8, (c) 3, and (d) 2 (black full lines). 
To average, $10^4$ waveguide realizations were used. The waveguides 
support $M=3$ open modes. Red dashed lines are best fittings of 
$\left\langle 1/T \right\rangle$ [$\left\langle \ln T \right\rangle$] with 
Eq.~(\ref{avR}) [Eq.~(\ref{avlnT})] for small [large] $L$.
The extracted $(\ell,\xi)$ are approximately equal to (a) $(44.77,134.38)$, 
(b) $(55.88,179.28)$, (c) $(94.97,265.78)$, and (d) $(136.88,346.61)$. 
Blue dot-dashed vertical lines indicate the positions of $\ell$ and $\xi$ 
which delimit the transport regimes: quasi-ballistic (I), diffusive (II), and 
localized (III). Red vertical lines mark the value of $M\ell$.}
\label{Fig2}
\end{figure*}

We remark that depending on the values of the parameters ($L_x$, $d$, 
$\epsilon$, and $N_T$), the classical (or ray) dynamics in each of the cavities can 
yield mixed or full chaos. However, here we consider only the case of full 
chaos \cite{MMLA}: $\epsilon d N_T/L_x^2 > 0.01$. Then, below we use $L_x=2\pi$, 
$d=L_x/2$, and $\epsilon=L_x/20$. This set of parameters produces full chaos 
for any $N_T\ge 1$. Finally, note that all lengths, here and below (including 
$\ell$, $\xi$ and $L$), are given in units of $L_x$.

We open the waveguide of length $L$ defined above by attaching two semi-infinite
collinear flat leads of width $d$ to its left and right ends. The leads support 
plane waves with energy $E = (\hbar^2/2\mathbf{m})[k_m^2 + (m\pi/d)^2]$,
where $k_m$ and $m\pi/d$ are, respectively, the longitudinal and transversal 
components of the total wave vector $K = \sqrt{ 2 \mathbf{m} E}/\hbar$.
Then, using finite element methods (see e.g. \cite{Smat1,Smat2}) we compute 
the scattering matrix, $S$-matrix, which has the form 
\[
S = \left(\begin{array}{cc}
      t & r' \\
      r & t' 
    \end{array} \right) \ ,
\]
where $t$, $t'$, $r$, and $r'$ are $M \times M$ transmission and reflection matrices.
Here, $M$ is the highest mode given by the largest $m$ beyond which the longitudinal 
wave vector $k_m=[2 \mathbf{m} E/\hbar^2 - (m\pi/d)^2]^{1/2}$ becomes complex.
Then, once the $S$-matrix is known we calculate the dimensionless conductance
from \cite{Landauer}
\begin{equation}
T = \mbox{Tr} (tt^\dagger) \ . 
\end{equation}
With this definition, the conductance can take values in the interval $[0,M]$.

The experimental realization of a scattering setup similar to ours has
been recently reported in Ref.~\cite{exp}. However, we notice that
in numerical simulations of transport through surface-disordered wires, step-like 
corrugated waveguides are more often used \cite{GTSN97,GS01,FGSMS02,GS05,QC,otros,FM11},
among others \cite{FYMS07,SFMY99}.
We also note that here we concentrate on the case of small number of open modes,
$M=[2,9]$; the case of $M\gg 1$ has been recently addressed in Ref.~\cite{FM11}.

\section{Results}

\subsection{Diffusive and localized regimes}

In order to identify the diffusive and localized transport regimes
in our corrugated waveguides, in Fig.~\ref{Fig2} we plot the average resistance 
$\langle 1/T \rangle$ and the average logarithm of conductance 
$\left\langle \ln T \right\rangle$ as 
a function of $L$ for waveguides with $N_T=18$, 8, 3, and 2.

For disordered wires, it is well established that 
(i) for relatively short wire lengths the resistance increases linearly
with $L$ as \cite{GTSN97,SFMY99,R1,R2}
\begin{equation}
\left\langle \frac{1}{T} \right\rangle = \frac{1}{M} + \frac{L}{M\ell} \ ;
\label{avR}
\end{equation}
while (ii) for relatively large wire lengths the conductance decays exponentially 
with $L$ in the form \cite{GTSN97,SFMY99,T1,T2}
\begin{equation}
\left\langle \ln T \right\rangle \propto - \frac{L}{\xi} \ .
\label{avlnT}
\end{equation}
We extract $\ell$ and $\xi$ from Fig.~\ref{Fig2} by performing fittings of the
data with Eqs.~(\ref{avR}) and (\ref{avlnT}), respectively, see red dashed lines.
In Fig.~\ref{Fig2} we also indicate, with blue dot-dashed vertical lines, the positions of 
the obtained $\ell$ and $\xi$. From this figure, it is clear that both $\ell$ and 
$\xi$ decrease by increasing $N_T$; equivalently, the mean-free path and the 
localization length increase when the disorder decreases.
Therefore, in Fig.~\ref{Fig2} we label as (II) and (III) the diffusive and
localized regimes, respectively. Additionally, we identify with (I) the quasi-ballistic 
regime, $L<\ell$, which we will not explore here.

It is also interesting to mention that we found that relation $\xi\approx M\ell$
(see for example \cite{GS01,GS05}) works well only when the surface corrugation 
is complex enough; i.e. for $N_T\ge 7$. See upper panels of Fig.~\ref{Fig2}. 
When $N_T\to 1$ we observe that $\xi< M\ell$. See lower panels of Fig.~\ref{Fig2}.
In any case the diffusive regime is clearly discernible in our calculations 
for all the values of $M$ and $N_T$ we used here. 

Note that to construct Fig.~\ref{Fig2} we have used waveguides supporting $M=3$
open modes (here and below, when showing results for different values of 
$M$ we always fix the energy such that $K=\pi(M+1/2)/d$).
We obtained similar plots for other values of $M$. However, by increasing
$M$ the values of $\ell$ and $\xi$ decrease and, as a consequence, the regimes 
(I) and (II) become narrower. For $M=9$, the highest $M$ we explored, the 
quasi-ballistic regime is hardly visible in the scale of Fig.~\ref{Fig2}.
In the following figures we will use values of $M$ different from 3 to emphasize 
that our results do not depend on the number of open modes, once the transport 
regimes are well determined.

\begin{figure}[t]
\includegraphics[width=7cm]{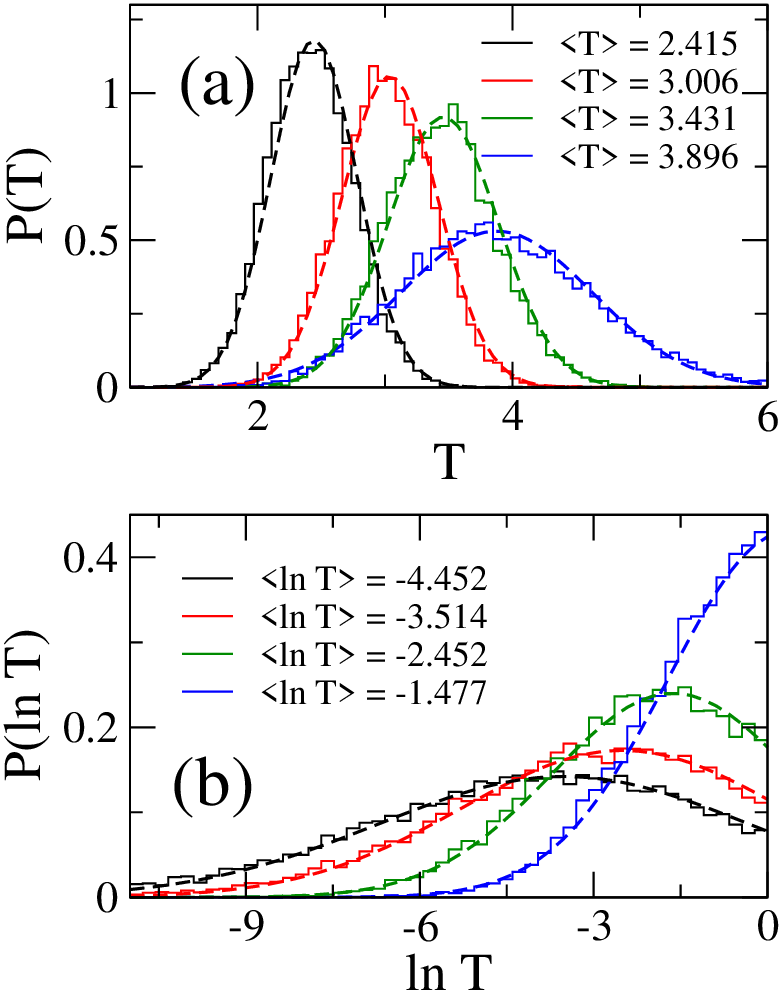}
\caption{(Color online) Conductance probability distributions (a) $P(T)$ and (b) 
$P(\ln T)$ for corrugated waveguides with $N_T=18$, 8, 3, and 2 (from left to right) 
in the (a) diffusive and (b) localized regimes. 
$10^4$ values of $T$ where used to construct each histogram.
The waveguides support $M=9$ open modes.
Dashed lines are (a) Gaussian and (b) log-normal distribution functions
characterized by the values of (a) $\bra T \ket$ and (b) $\bra \ln T \ket$ 
reported in the corresponding panels.}
\label{Fig3}
\end{figure}

Then, in Fig.~\ref{Fig3}, we verify the predictions for the conductance probability 
distribution function in the diffusive and localized regimes. 
As clearly shown in this figure, where $M=9$ open modes were considered, $P(T)$ has 
a well defined Gaussian shape in the diffusive regime [Fig.~\ref{Fig3}(a)], while 
$P(\ln T)$ has the log-normal form in the localized regime [Fig.~\ref{Fig3}(b)].
We know that the results reported in Fig.~\ref{Fig3} are already expected and 
thus it may seem unnecessary to shown them. However, we decided to present them in
order to stress that the transport properties, exemplified here by the form of $P(T)$, 
of rough and smooth corrugated waveguides are in fact similar except at the crossover 
regime, as will be shown below. 

\subsection{Crossover regime}

\begin{figure*}[t]
\includegraphics[width=12cm]{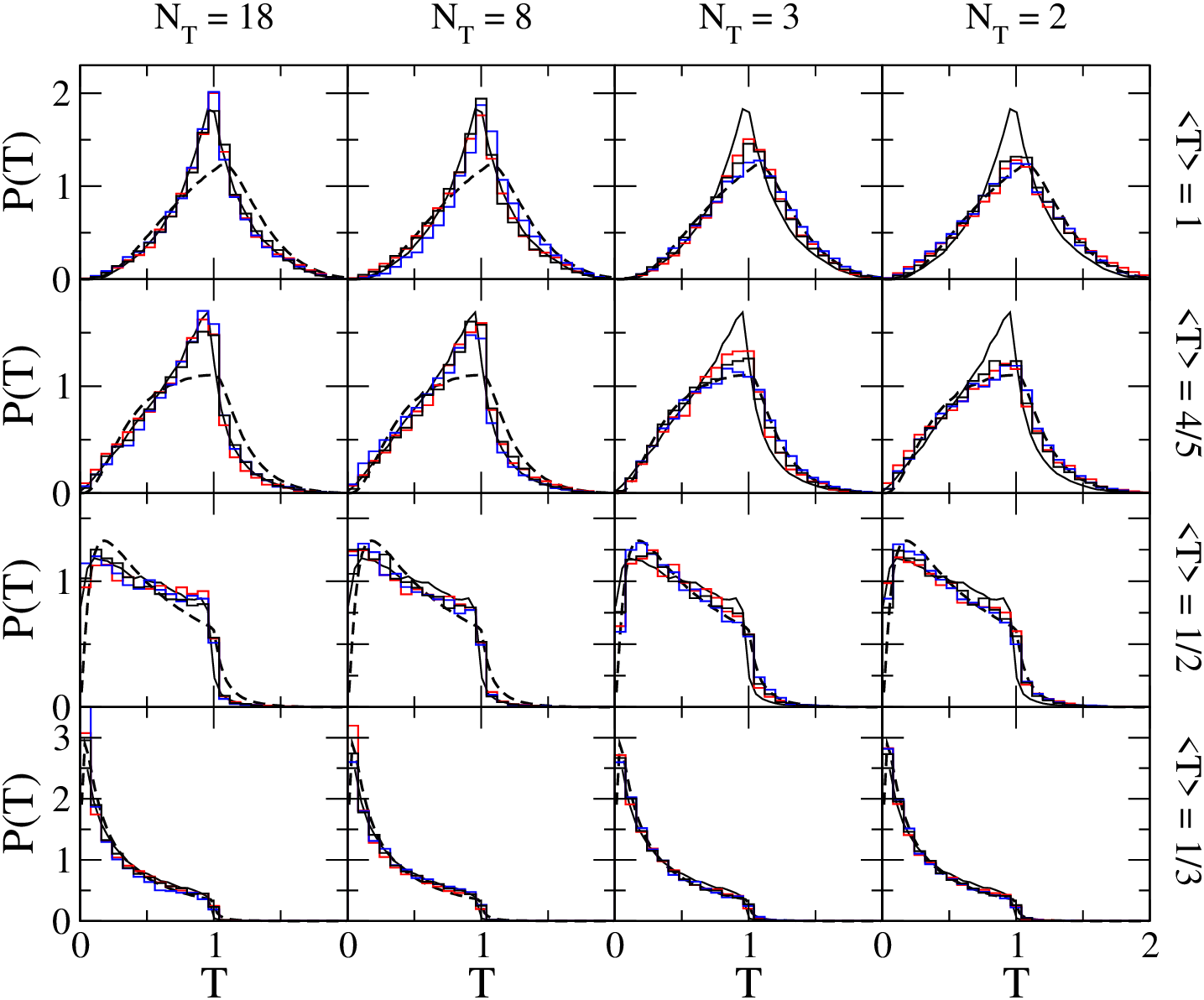}
\caption{(Color online) Conductance probability distributions $P(T)$ for corrugated 
waveguides with $N_T=18$, 8, 3, and 2 (columns) for $\bra T \ket=1$, 4/5, 1/2, and 
1/3 (rows). Each panel contains three histograms corresponding to waveguides with
$M=5$, 7 and 8. $10^4$ values of $T$ where used to construct each histogram.
Continuous and dashed lines are the surface-disorder FYMS and the bulk-disorder 
DMPK predictions for $P(T)$, taken from \cite{FYMS07} and 
\cite{FGSMS02}, respectively.}
\label{Fig4} 
\end{figure*}

As already anticipated, we have found that the most interesting result for 
the distribution of conductances appears at the crossover regime,
where $\bra T \ket \sim 1$. Here, we observe that the shape of $P(T)$ does depend 
on the waveguide corrugation complexity. To show this, in Fig.~\ref{Fig4} we plot
$P(T)$ for corrugated waveguides with $N_T=18$, 8, 3, and 2 (columns) for some
values of $\bra T \ket$ (for comparison purposes in Fig.~\ref{Fig4} we chose the 
same values of $\bra T \ket$ used in \cite{FYMS07,GTSN97,GS01,FGSMS02,GS05}).
As a reference, we also include the surface-disorder FYMS and the bulk-disorder 
DMPK predictions for $P(T)$.
In each panel we plot three histograms corresponding to waveguides supporting 
$M=5$, 7, and 8 open modes.
Note that for large $N_T$, in our case $N_T=18$ \cite{note4}, the shapes of
the numerically obtained $P(T)$ are well described by the FYMS prediction 
for surface-disordered waveguides, as expected (see panels in the 
left-most column of Fig.~\ref{Fig4}). However, once the waveguide corrugation 
complexity is 
decreased, important deviations appear. Moreover, when $N_T=2$, $P(T)$ fully 
coincides with the DMPK prediction for bulk-disordered waveguides (see panels in 
the right-most column of Fig.~\ref{Fig4}). This fact is
more evident for $\bra T \ket=1$ and 4/5 where the differences between FYMS 
and DMPK predictions are easily distinguishable.

So, we observe an effective and smooth evolution in the form of $P(T)$, from 
the surface-disorder FYMS to the bulk-disorder DMPK predictions, as a function 
of the (decreasing) waveguide corrugation complexity $N_T$. 
We understand this result in the following simple way.
In the limit $N_T\to 1$ our waveguide can be considered as a linear chain
of coupled chaotic cavities, each one defined by the cosine billiard \cite{linda}.
Then, the model of quantum dots in series of Iida, Weidenmuller, and Zuk 
applies \cite{IWZ90}. Moreover, this model reduces to a supersymmetric nonlinear 
$\sigma$ model \cite{EL83,E83,MMZ94} which turns out to be equivalent to the DMPK 
approach \cite{BF96}. That is, while FYMS describes well the limit $N_T\gg 1$ of
our corrugated waveguide, DMPK should describe the limit $N_T\to 1$; as we 
in fact observe in Fig.~\ref{Fig4}.
Therefore, our waveguide model shows a disorder-to-chaos transition in the 
shape of the conductance distribution.

Finally, we want to add that the parameters used in 
Refs.~\cite{GTSN97,GS01,FGSMS02,GS05}, translated to our symbols,
are \cite{note5}: $d/\epsilon = 7-13.25$ (here we used $d/\epsilon = 10$) and 
$N_T\approx 7$. Then, since we are observing well developed 
surface-disorder FYMS transport properties in our corrugated waveguides 
for $N_T\ge 8$, there is no contradiction between our results and those 
presented in Refs.~\cite{GTSN97,GS01,FGSMS02,GS05}.
Moreover, we think that it would be interesting to explore the limit where the step length,
in the step-like corrugated waveguide model of Refs.~\cite{GTSN97,GS01,FGSMS02,GS05},
becomes of the order of the waveguide width; which is somehow equivalent to the limit 
$N_T\to 1$ in our waveguide model.

\section{Conclusions}

We have studied the distribution of conductances $P(T)$
for a quasi-one-dimensional corrugated waveguides with tunable   
corrugation complexity: from rough, $N_T=18$, to smooth, $N_T=2$. 
We verified that both, the mean free path and the localization length decrease
for increasing $N_T$. Also, we confirmed that $P(T)$ and $P(\ln T)$ have the 
Gaussian and log-normal shapes in the diffusive ($\bra T \ket> 1$) and localized 
($\bra T \ket\ll 1$) transport regimes, respectively. 

At the crossover between the diffusive and the localized regime, 
$\bra T \ket \sim 1$, we reported that $P(T)$ monotonously evolves from the 
surface-disorder FYMS (when $N_T=18$) to the bulk-disorder DMPK (when $N_T=2$) 
predictions for decreasing $N_T$. 
We understood this behavior as a consequence of the underlying deterministic 
dynamical chaos; since the waveguides having smooth boundaries (i.e.,
when $N_T\to 1$) are effectively linear chains of attached chaotic 
cavities.

We believe that our results, as well as our model of corrugated wires with tunable 
corrugation complexity, may stimulate further analytical and numerical studies on 
the transport properties at the diffusive-to-localized transition regime.

\begin{acknowledgments}
We acknowledge support form VIEP-BUAP grant MEBJ-EXC12-G
and Fondo Institucional PIFCA 2012 (BUAP-CA-169).
\end{acknowledgments}


\begin{thebibliography}{99}
\bibitem{note4}
We have verified that our conclusions do not change when considering 
larger values of $N_T$.
\bibitem{note5}
Evidently, the parameter $N_T$ does not exist in the step-like 
corrugated waveguide model of 
Refs.~\cite{GTSN97,GS01,FGSMS02,GS05}.
However, there the step length $l$ plays the role of the inverse 
of $N_T$ in our waveguide model. In fact, by discretizing our 
smooth cosine-like corrugation we could write $l\equiv L_x/2N_T$.
\end{thebibliography}
\end{document}